\RequirePackage{fix-cm}
\documentclass[smallextended]{svjour3}       
\smartqed  
\usepackage{lineno,hyperref}
\usepackage[utf8]{inputenc}
\usepackage[english]{babel}
\usepackage{amsmath, latexsym}
\usepackage{amsfonts}
\usepackage{afterpage}
\usepackage{amssymb}
\usepackage{graphicx}
\usepackage{subcaption}
\usepackage{float}
\usepackage{subcaption}
\usepackage{setspace}
\usepackage{array}
\usepackage{lscape}
\usepackage{natbib}
\begin{document}

\title{\textit{Graph\_sampler}: a simple tool for fully Bayesian analyses of
DAG-models }



\author{Sagnik Datta         \and
        Ghislaine Gayraud      \and
        Eric~ Leclerc                \and
        Frederic Y. Bois 
}


\institute{Sagnik Datta \and  Ghislaine Gayraud\at
            Sorbonne Universités, Université de Technologie de Compiègne, 
BMBI (S. Datta), LMAC (G. Gayraud)
 Bât. G.I., CS 60 319
60 203 Compiègne cedex, France\\
              Tel.:  +33 3 44 23 47 37 and
              Fax:  +33 3 44 23 44 77\\
              \email{sagnik.datta@utc.fr}  and \email{ggayraud@utc.fr}
           \and
              Eric Leclerc\at
              LIMMS/CNRS-IIS (UMI 2820)
Institute of Industrial Science, The University of Tokyo, 4-6-1 Komaba, Meguro-ku, Tokyo 153-8505, Japan\\
             \email{eleclerc@iis.u-tokyo.ac.jp}     
             \and
            Frederic Y. Bois\at
             INERIS
             Parc ALATA, BP2
             60550, Verneuil en Halatte
             France
}

\date{Received: date / Accepted: date}

\maketitle

\begin{abstract}
Bayesian networks (BNs) are widely used graphical models usable to draw statistical inference about Directed acyclic graphs (DAGs). We presented here \textit{Graph\_sampler} a fast free C language software for structural inference on BNs. \textit{Graph\_sampler} uses a fully Bayesian approach in which the marginal likelihood of the data and prior information about the network structure are considered. 
This new software can handle both the continuous as well discrete data and based on the data type two different models are formulated. The software also provides a wide variety of structure priors which can be informative or uninformative.
We proposed a new and much faster  jumping kernel strategy in the Metropolis-Hastings algorithm. 
The source C code distributed is very compact, fast, uses low memory and disk storage. We performed out several analyses based on different simulated data sets and synthetic as well as real networks to discuss the performance of \textit{Graph\_sampler}. 
 
\section*{Acknowledgment} S.Datta is funded by a Ph.D. studentship for the French Ministry of Research.
\end{abstract}

\section{Introduction}
Representing knowledge with uncertainty and automatic reasoning is often carried out using graphical models \citep{Pearl88,Lau96,Nep89}. Judea Pearl and Richard E. Neapolitan were the first to summarize the properties of directed acyclic graphs (DAGs) and established them as a new field of study. In the recent years, formal statistical inference on systems of multiple interacting components is often done using DAGs \citep{Hec95} or Markov networks \citep{Ed00}. A Bayesian network (BN) or a belief network is a probabilistic model denoted by a graph $\mathcal{G} = ( \mathcal{V,E})$ in which each node or vertex $v$ $\in \mathcal{V}$ represents one of the random variables in set X = $(X_1, X_2,...,X_N)$, where N is the number of nodes and each edge e $ \in \mathcal{E}$ express the dependence among the variables in X. A BN is always directed and acyclic and is therefore a DAG. Besides static BNs, there are also dynamic BNs \citep{Hus03,Fri98}, which are actually  generalizations of hidden Markov models. In that paradigm, all the random variables of the network are considered to be potentially related to each other over adjacent time points.

DAGs have been used for more than two decades in biomedicine and health-care for handling uncertainty in disease diagnosis, selecting the optimal treatment and predicting treatment outcome \citep{And92}. Other applications are found in social network analyses, high dimensional data analyses etc. In computational biology and bioinformatics, BNs have been proposed to model DAGs, for example, for modeling gene regulatory networks, protein structure and gene expression. Many of the available software \citep{Kor10} programs which do BN parameter estimation or structural inference run on commercial platforms. The others which are readily available and free are generally not well maintained and not updated regularly. However there are some widely used packages scripted in R that are free and properly maintained. 

In this article we present an efficient C language software \textit{Graph\_sampler} for Bayesian inference on the structure of BNs; this latter is based on the well-known  Metropolis-Hastings(MH) algorithm that allows to sample DAGs from the posterior distribution. Unlike the existing BN learning software \textit{Graph\_sampler} propose a new jumping kernel in the MH algorithm making it more time efficient. The software is easy to use and works well with both discrete and continuous data. For each of these data types we formulated two different models. One of the key feature of this software is that it handles large network structure efficiently. A number of different priors are included in this software to reflect the prior knowledge on the graph structure. This software is quite flexible for using different combination of prior knowledge. We made several performance test on our software with intensive simulation studies regarding all possible choices of models and priors to venture the robustness of \textit{Graph\_sampler}. Alongside we considered \textit{structmcmc}, a versatile and easy to use R package \citep{Muk08} for detailed discussion on the performance of \textit{Graph\_sampler}. Like \textit{Graph\_sampler}, this software also aims to learn structure on DAGs via a fully Bayesian approach.
 
The remainder of the paper is organized as follows: Section 2 briefly discusses the statistical inference for BNs with \textit{Graph\_sampler}; Section 3 is about the new jumping kernel, the installation and running of \textit{Graph\_sampler}; Section 4 discusses about some of the widely used software and their approach. This section also describes the methodology to check the efficiency of \textit{Graph\_sampler} for discrete as well as continuous data set along with the difficulties in the flipping technique. In Section 5 we present our results along with their significance. In Section 6 we summarize all our results in the form of a discussion. 
 
\section{Statistical inference for BNs} 

Recent work on BNs by Mukherjee and Speed \citep{Muk08}, used Markov chain Monte Carlo (MCMC) simulations to infer on network structure from node values. They also considered priors encoding information relative to existence of edges, degree distribution and sparsity structure of the graph. \cite{Boi13} recently extended informative priors to account for motifs frequencies in order to generate realistic gene regulating graphs \citep{Boi13}. For baseline information they proposed to use Bernoulli distributions to model prior knowledge on individual edges. 

Variational Bayes (VB) methods have also been proposed as an alternative to full Bayesian inference. VB is an efficient way to deal with intractable integrals arising in a full Bayesian context and can be considered as an entension of the expectation-maximization (EM) algorithm \citep{Mat03}. VB can also help in model selection by providing a lower bound for the data marginal likelihood. However it provides only an approximate analytical solution for the posterior probability of the parameters and latent variables involved in a graphical model. Hence we focus here on a full Bayesian approach.

The following sections describe the prior densities and data likelihood available in \textit{Graph\_sampler}.

\subsection{Priors on graph structure}

It is quite a challenging problem to make inference on graphical model structure even with moderate number of nodes, partly because of the huge number of possible graphs. For $N$ nodes, the number of possible DAGs can be computed recursively as \citep{Rob73}:
\begin{eqnarray}
a_N &=& \sum\limits_{k = 1}^N (-1)^{k-1} \binom Nk 2^{k(N-k)} a_{N-k}\nonumber
\end{eqnarray}
where $a_0 = 1$ by convention. So for 2, 3, 4 and 5 nodes network there are 3, 25, 543 and 29281 possible DAGs respectively. 

All of those possible graphs are usually not equally plausible a priori
and thus certain features may be incorporated as more likely than the others, a priori. To make inference on graph structure, Bois and Gayraud \citep{Boi13} considered three different priors. In the most general case, a set of independent Bernoulli priors ($P_B$) is used to model the prior knowledge on individual edges. If $p_{i,j}$ is the probability of existence of a directed edge $e_{i,j}$ from 
node $\textit{i}$ to node $\textit{j}$, and $e_{i,j}  \sim B(p_{i,j})$ for all $(\textit{i, j}) \in \{1,...,N\} \times \{1,...,N\}$, then the global Bernoulli prior $P_B$ on the graph $G$ is
\begin{eqnarray}
P_{B}(G) &=& \prod\limits_{i \neq j=1}^N (p_{i,j})^{e_{i,j}} (1 - p_{i,j})^{1- e_{i,j}}\nonumber
\end{eqnarray}
The choice of the value for each hyperparameter $p_{i,j}$ depends on the prior evidence we have on the existence of the given edge from the scientific literature.

The degree $deg(v)$ of a vertex $v$ is defined as the number of edges involving $v$. We can also define the degree distribution $(P_D)$ for $G$ as a function $P_d = card\{v \in \mathcal{V} : deg(v) = d\}$ for all the vertices having degree $d$. The prior on degree can be expressed as a power law given by
\begin{eqnarray*}
P_d \propto d^{-\gamma}, \qquad\text{with}\qquad \gamma > 0 
\end{eqnarray*}
Thus we can define the prior degree distribution for the graph $G$ as 

\begin{eqnarray*}
P_{D}(G) \propto \prod \limits_{i=1}^N \left(\sum \limits_{j = 1}^{N} e_{i,j}\right)^{-\gamma}, \qquad\text{with}\qquad \sum \limits_{j = 1}^{N} e_{i,j} > 0 
\end{eqnarray*}
In addition to $P_B$ and $P_D$, if we consider a Beta-Binomial prior ($P_M$) as it is done in \cite{Boi13} on the occurence of three-nodes motifs then the total prior on the graph $G$ can be expressed in a product form as:
\begin{eqnarray}
P_{T}(G) &\propto & P_{B}(G) \times P_{D}(G) \times P_{M}(G)\nonumber
\end{eqnarray}

Alternative to the Bernoulli prior for the presence of edges is the so called concordance prior ($P_C$) (see \cite{Muk08}). The latter required the specification of a prior matrix E with elements $E_{i,j} = 1$ representing a desired edge and -1 representing a non-desired edge. At each iteration the prior is calculated by counting the number of disagreements with the adjacency matrix \textit{A} with elements $A_{i,j} = 1$ $or$ $0$ representing the presence or absence of a directed edge starting from node \textit{i} and ending in node \textit{j} of  the graph G. The form of the concordance prior is then
\begin{eqnarray}
P_{C}(G) &\propto & \exp( - \rho( \sum \limits_{i,j=1}^{N}\vert{A_{i,j} - E_{i,j}} \vert))\nonumber
\end{eqnarray}
where $\rho$ is a positive valued hyperparameter.
If $P_C$ is used, then the total prior is:
\begin{eqnarray}
P_{T}(G) &\propto & P_{C}(G) \times P_{D}(G) \times P_{M}(G) \nonumber
\end{eqnarray}
or, equivalently, if $P_B$ is a flat prior (with $p_{i,j}$ = 0.5 for $i \neq j$) to avoid any conflict or double accounting with $P_C$, the total prior is:
\begin{eqnarray}
P_{T}(G) &\propto & P_{B}(G) \times P_{C}(G) \times P_{D}(G) \times P_{M}(G) \label{tot}
\end{eqnarray}

\subsection{Data likelihood and prior predictive distribution}

Our main interest is to uncover the underlying structure of BNs. In any Bayesian network, a parent node has always an influence on its child nodes. Let us denote by $x$ = $(x_1,...,x_N)$ the data we have on N nodes, where $x_i$ is a \textit{n}-dimensional vector; where $n$ is the number of data points per node. Even though the model considered involves many parameters, the posterior distribution of each of these parameters are not of our primary interest. Integrating out the parameters in a Bayesian context leads to the prior predictive distribution (or the joint marginal likelihood of the data): $f(x \vert G)$ = $\prod\limits_{i=1}^N f(x_i \vert P_a(x_i))$, where $P_a(x_i)$ is the set of parent values of $x_i$ in graph G \citep{Hec95} and $f( \cdot )$ is the prior predictive distribution of $x_i$ given its parenthood. For a global parent $P_a(x_i)$ = $\emptyset$ and thus $ f(x_i \vert P_a(x_i))$ reduces to $f(x_i)$.

\subsubsection{Continuous data}
\subsubsection*{\textbf{Gaussian regression model with a Normal-Gamma prior}}
       The general expression for the linear Gausian regression for a given node $x_i$ is:
\begin{eqnarray}
x_i &=& M(x_i)\beta + u \label{reg}
\end{eqnarray}
where $x_i =(x_{1,i},x_{2,i},\cdots,x_{n,i})$ is a vector of \textit{n} observations of the dependent variable
$x = ((x_{i,j})_{1 \leq i \leq n;1 \leq j \leq k})$, $k$ is the cardinality of $P_a(x_i)$,  $M(x_i)$ is a so-called design matrix of order \textit{(n $ \times $ (k+1))} with the first column as 1's and other columns as $P_a(x_i)$, $\beta$ is a real valued vector of regression parameters of length k, and  u follows a Gaussian distribution ${\cal N} (0, \lambda^{-1}I_n)$ distribution with $\lambda$ being a positive real valued precision and $I_n$ is the identity matrix of size $n$. The likelihood for this regression model is therefore:

\begin{eqnarray}
L (\lambda , \beta \vert  x_i, P_a(x_i)) = (\frac{\lambda}{2 \pi})^{n/2} \; \exp ( -
\frac{\lambda}{2} (x_i-M(x_i) \beta)^{t}(x_i-M(x_i) \beta) )  \nonumber
\end{eqnarray}
where $C^{t}$ denotes the transpose of the matrix \textit{C}.

It is classical to choose conjugate form for the priors on the parameters (i.e. $\beta$ and $\lambda$) involved in the regression model :
\begin{eqnarray}
P(\beta , \lambda) &=&N_k(\beta\mid\beta_0,(n_0\lambda)^{-1})Ga(\lambda\mid\alpha,\omega)\nonumber\\
&=&\frac{(\omega)^{\alpha}(n_0)^{k/2}}{(2\pi)^{k/2}\Gamma(\alpha)}(\lambda)^{\frac{k}{2}+\alpha-1}\exp\left(-\frac{\lambda}{2}(\beta-\beta_0)^{t} n_0 (\beta-\beta_0)-\omega\lambda\right) \nonumber
\end{eqnarray}
where $\beta_0$ (real valued vector) and $n_0$ (matrix of dimension $k$$\times$$k$) are hyper-parameters related to $\beta $ and $ \alpha , \omega $, both being real valued positive numbers, are hyper-parameters of $\lambda $. In the above equation $\Gamma(.)$ represents the Gamma function.

In that case, the prior predictive distribution is $t_\nu(\mu,\Sigma)$, the n-dimensional multivariate t-distribution with parameters $\mu,\Sigma$ and $\nu$, whose density function is:
\begin{eqnarray}
f(x_i\vert P_a(x_i))&=&\frac{\Gamma(\nu+n)/2}{\Gamma(\nu/2)(\nu\pi)^{n/2}}\mid\Sigma\mid^{-1/2}\left[ 1+\frac{1}{\nu}(x_i-\mu)^{t}\Sigma^{-1}(x_i-\mu)\right] ^{-\frac{\nu+n}{2}} \label{nor}
\end{eqnarray}
where, $\mu= [\mu_1,…, \mu_n]^{t}$ is the location parameter, $\Sigma$ is the scalar matrix of dimension ($k \times k$) and $\nu$ is the degrees of freedom such that
\begin{eqnarray}
\mu&=&M(x_i)\beta_0\nonumber\\
\Sigma^{-1}&=&h(M(x_i))\alpha\omega^{-1} \nonumber\\
h(M(x_i))&=&I_n-M(x_i)[M(x_i)^{t}M(x_i)+n_0]^{-1}M(x_i)^{t} \nonumber\\
\nu&=&2\alpha\nonumber
\end{eqnarray}

\subsubsection*{\textbf{Gaussian regression model with a Zellner g-prior}}
With \textit{Graph\_sampler} we can also use the Zellner g-prior \citep{Muk08,Not04} for the parameters $\beta$ and $(\lambda)^{-1}$:
\begin{eqnarray}
P(\beta \vert \lambda ^{-1}) &= &  N_k (0, g\lambda^{-1}[M(x_i)^{t}M(x_i)]^{-1}) ,\nonumber\\
P(\lambda^{-1}) & \propto & \lambda \nonumber
\end{eqnarray}
where \textit{g} is a user defined positive scale factor. The prior predictive distribution of the data is given by:
\begin{eqnarray}
f(x_i \vert P_a(x_i)) &\propto& (1 + g)^{-(k + 1)/2} s^{-n/2} \label{zel}
\end{eqnarray}
where \textit{k} is the cardinality of $P_a(x_i)$ and 
\begin{eqnarray}
s &=& x_i'x_i - \frac{g}{1+g} x_i^{t} M(x_i) [M(x_i)^{t}M(x_i)]^{-1} M(x_i)^{t} x_i \nonumber
\end{eqnarray}
provided the term $M(x_i)^{t}M(x_i)$ is invertible. A sufficient condition for $M(x_i)^{t}M(x_i)$ to be invertible is that $k \leqslant n$, that is, the number of parents should be less than the number of data points for per node. Thus Zellner g-prior fails to work when number of parents is greater than the number of data points per node.

\subsubsection{Discrete data}
For discrete data, \textit{Graph\_sampler} offers the possibility to use a Multinomial model with a Dirichlet prior on its parameters (see \citep{Hec95}). For such a prior on the parameters we have a closed form of the prior predictive:
\begin{eqnarray}
f(x \vert G) &=& \prod \limits_{i=1}^n \prod \limits_{j=1}^{s_i} \frac{\Gamma(D'_{ij})}{\Gamma(D'_{ij} + D_{ij})} \cdot \prod \limits_{k=1}^{m_i} \frac{\Gamma(D'_{ijk} + D_{ij})}{\Gamma(D_{ijk})} \label{dir}
\end{eqnarray}
where $D_{ijk}$ is the number of components of $x_i$ that takes the value \textit{k} given that $P_a(x_i)$ has configuration $j$, $D'_{ijk}$ are the Dirichlet hyperparameters, $s_i$ represents the possible number of configurations of the parents of $x_i$,  $m_i$ stands for the number of possible values of components of $x_i$ and $D'_{ij}$ and $D_{ij}$ are given by:

\begin{equation}
D'_{ij} = \sum \limits_{k=1}^{m_i} D'_{ijk} \qquad\text{and}\qquad  D_{ij} = \sum \limits_{k=1}^{m_i} D_{ijk}. \nonumber
\end{equation}

 \section{\textit{Graph\_sampler}: sampling and installation}
\subsection{Efficient sampling: fast jumping kernel} 
 \textit{Graph\_sampler} can efficiently generate random samples for general directed graphs \citep{Boi13}, but we focus here on the sampling of BNs from a posterior distribution conditionned by data (observed node values). We use an adjacency matrix representation for the graph and store only the eventual difference between adjacency matrix as it is a fast and efficient storage method.
A Metropolis-Hasting sampler \citep{Cas04} is used to sample random graphs according to the prescribed posterior probability distribution.
The algorithm of the simplified jumping kernel is as follows for the $t-th$ iteration:

We denote the current graph by $G$ and  its adjacency  matrix by $A^{G^t}$. The proposal graph is denoted by $G'$ and its adjacency matrix by $A^{G^{'}}$ in our algorithm.

\subsubsection*{\textbf{Algorithm}}

Step 1: Select $A_{i,j}^{G^{t}}$ while scanning $A^{G^t}$ 

\noindent Step 2:

   \begin{itemize}
     \item[(a):]  Sample $ z_{i,j} \sim  Bernoulli (p_{i,j})$ where $p_{i,j}$ is the Bernoulli prior for edge $e_{i,j}$
     \item[(b):]
         \begin{itemize}
           \item[(i):] \textit{Adding an edge}\\
                            if $z_{i,j} =1$, then 
$A_{i,j}^{G^{'}} =\left\{ \begin{array}{ccl}
A_{i,j}^{G^{t}} & \mbox{{\rm if}} & A_{i,j}^{G^{t}}=1\\
1 & \mbox{{\rm o.w. and }} &  \mbox{ {\rm provided $G'$ is still a DAG, }} \\
&&  \mbox{ {\rm o.w. go back to Step 1}} 
 \end{array}\right.$ 

             \item[(i):] \textit{Deleting an edge}\\    
                             if $z_{i,j} =0$, then 
$A_{i,j}^{G^{'}} =\left\{ \begin{array}{ccl}
A_{i,j}^{G^{t}} & \mbox{{\rm if}} & A_{i,j}^{G^{t}}=0\\
0 & \mbox{{\rm if}} & A_{i,j}^{G^{t}}=1\\
 \end{array}\right.$ 
           \end{itemize}
     \end{itemize}

\noindent
Step 3: Calculate the acceptance ratio. We accept $G'$ with probability\\
\begin{eqnarray}
\delta = \min ( 1, \frac{f(x,G')P_{T}(G')P(A_{i,j}^{G^{t}}\vert A_{i,j}^{G^{'}})} {f(x,G^t)P_{T}(G^t)P(A_{i,j}^{G^{'}}\vert A_{i,j}^{G^{t}})}) \nonumber
\end{eqnarray}
\indent
Note that due to Step 2, the acceptance ratio can be rewritten as follows:
\begin{eqnarray}
\delta = \min ( 1, \frac{f( x\vert G') P_{T}(G') P_{B}(G^t)} {f( x \vert G^t) P_{T}(G^t) P_{B}(G')})\nonumber
\end{eqnarray}
where $f$($x$$\vert$$G$) =  $\prod\limits_{i=1}^N f(x_i \vert P_a(x_i))$ is the prior predictive given by Eq (\ref{nor}, \ref{zel} and \ref{dir}) and $P_{T}$ is the total prior on the graph structure. Clearly this simplifies since $P_B$ is a part of $P_T$ (see Eq(\ref{tot})).\\
\noindent
Step 4: Choose $G^{t+1}$ as follows:
$G^{t+1} =\left\{ \begin{array}{ccl}
G^{'} & \mbox{{\rm with}} & \mbox{{\rm probability  }}\delta\\
G^{t} & \mbox{{\rm with}} & \mbox{{\rm probability  1-}} \delta\\
 \end{array}\right.$ 

\noindent

The procedure is repeated until convergence in probability is attained. Gelman and Rubin's (GR) $\hat{R}$
criterion \citep{Gel92} is used on each element of the graph's adjacency matrix to check the convergence of several simulation chains. The advantage of using the GR criterion for the convergence is that, we do not have to save the whole chains from the start. We can check the convergence using the final posterior edge probability matrix obtained after the specified number of iterations neglecting the burning runs. For this criterion, we consider the three edge probability matrices and calculate the within-chain and the between-chain variance. Then the estimated variance of the parameter is calculated as a weighted sum of the within-chain and between-chain variance. Based on the potential scale reduction factor we infer on the convergence of the three chains.
For BNs we need to ensure that the proposed graphs are DAGs. This is done with a fast topological sorting algorithm (similar to that of \citep{Per06}) operating on a list index of the nodes.

\subsection{Graph\_sampler installation}

\textit{Graph\_sampler} is an easily available free software that can be redistributed or modified under the terms of the GNU General Public License as published by the Free Software Foundation. It is an inference as well as simulation tool for DAGs and can simulate random graphs for general directed graphs as well as for DAGs. In the case of BNs, we infer about their probable structure through the joint use of priors and data about node values.

\textit{Graph\_sampler} is written in ANSI-standard C language and can be compiled in any system having a ANSI C compliant compiler. The GNU gcc compiler (freeware) is highly recommended and the automated compilation script (called Makefile) can be successfully used if the standard 'make' command is available. In order to modify the input file parser, the 'lex' and ' yacc' are highly recommended. The full software along with the manual can be downloaded from:

https://sites.google.com/site/utcchairmmbsptp/software

Once downloaded, the software should be decompressed using 'gunzip' and 'tar' commands. Other archiving tools can also be used. \textit{Graph\_sampler} can be compiled using the 'make' command. On successful compilation of \textit{Graph\_sampler}, it is ready for running. In order to run \textit{Graph\_sampler}, an input file specifying the simulation parameters should be provided. In Unix the command-line syntax to run that executable is:

"\textit{graph\_sampler [input-file [output-prefix]]}"

where the brackets indicate optional arguments. If no input file and/or output prefix are not specified, the program uses the defaults. The default input file is \textit{script.txt} and the output files created depends on the parameters specified in the input file. Default output file names are \textit{best\_graph.out, graph\_samples.out, degree\_count.out, motifs\_count.out,}
\textit{edge\_p.out} and\\
\textit{results\_mcmc.bin}.

A \textit{Graph\_sampler} input file is a text (ASCII) file that obeys relatively simple syntax (see the manual). Values of all the predefined variables in the input file should be properly defined. Description and range of each variable is illustrated in the manual. In case of improper assignment of values, \textit{Graph\_sampler} post error messages during runtime.

\section{Efficiency analyses of \textit{Graph\_sampler}}

In order to evaluate the efficiency and accuracy of \textit{Graph\_sampler}, we performed several experimental runs based on different network structure, size and data type. We even varied the underlying model based on the data type to have a clear idea about the efficiency of the software. There are other very well known packages in R like the \textit{deal} and \textit{bnlearn}. The package \textit{deal} \citep{Sus03} is scripted in R language and uses BNs to analyse the data which can be discrete and/or continuous types; for the network parameters, suitable priors can be constructed and parameter estimation is possible using successive updating. This package is useful for structure learning of the network and uses the heuristic greedy search algorithm. The scoring function is based on maximising the Bayes factor. At each step of the greedy search algorithm, we either add, delete or reverse an edge and calculate the Bayes factor for all the possible graphs. The proposal graph with the maximum Bayes factor is selected to update the current graph. The package \textit{bnlearn} \citep{Scr10} is also scripted in R to do structural inference on BNs. This package is efficient to work with both discrete as well as continuous data. For the BN structure learning, various constraint-based algorithms like the \textit{Grow-shrink}, \textit{Incremental association} and \textit{Max-min parents and children} algorithms are implemented. \textit{Hill-Climbing} greedy search algorithm is the only score-based algorithm implemented in this package. The sampling algorithm consists of first recovering the skeleton of the desired graph and then evaluating the directionality of the edges. Both of these packages sample the best graph that maximizes the Bayes factor or the posterior odds from amongst all possible proposal graphs. This is quite different from the sampling scheme that we proposed. Our interest is to sample graphs based on the specified posterior distribution. Another efficient Matlab package for BN analysis is \textit{bnt} \citep{Murphy07}. This tool does inference on parameters as well as on network structure. This software is suitable for both decision networks and dynamic Bayesian networks. However this tool lacks the GUI. Inference is carried out using various algorithms like the Junction tree, variable elimination and Pearl's polytree. \textit{bnt} has various options for parameter learning as well as for structure learning. It is very clearly documented, free and is very object-oriented. Besides these well known packages, we also have \textit{structmcmc} \citep{Muk08} which is an efficient software coded in R for BN structure learning. We selected \textit{structmcmc} because of its similarities with \textit{Graph\_sampler} regarding the model formulation and the Metropolis-Hasting algorithm. Unlike \textit{Graph\_sampler}, the sampling technique in \textit{structmcmc} also allows flipping of edges between two selected nodes. This package also propose to sample graphs based on the posterior distribution. Thus we selected \textit{structmcmc} as a baseline software to discuss in details the advantages and limitations of \textit{Graph\_sampler}. \textit{structmcmc} proposes a Zellner g-prior for continuous datasets and a Multivariate Dirichlet prior for discrete datasets. The Normal Gamma model is not available to \textit{structmcmc} users.

In all the cases we used a null matrix as the initial adjacency matrix. Results showed that they are robust with respect to the initial adjacency matrix. For the prior on edges, we used the concordance prior \citep{Muk08} $(P_C)$ with $\rho = 1$ since \textit{structmcmc} does not provide a Bernoulli prior. The prior on the loop motifs $P_M$ was not used and therefore set to 1. We also used the degree prior $(P_D)$ to check the increase in efficiency of \textit{Graph\_sampler}. As an alternative to $(P_C)$, we also used both informative as well as uninformative $(P_B)$ prior as a structure prior to check the efficiency of the software. We followed the simulation procedure described in \citep{Muk08} to generate discrete datasets. For the continuous case, we generate data as described in Equation (\ref{reg}).

\begin{figure}[H]
\centering
\includegraphics[width=0.8\linewidth]{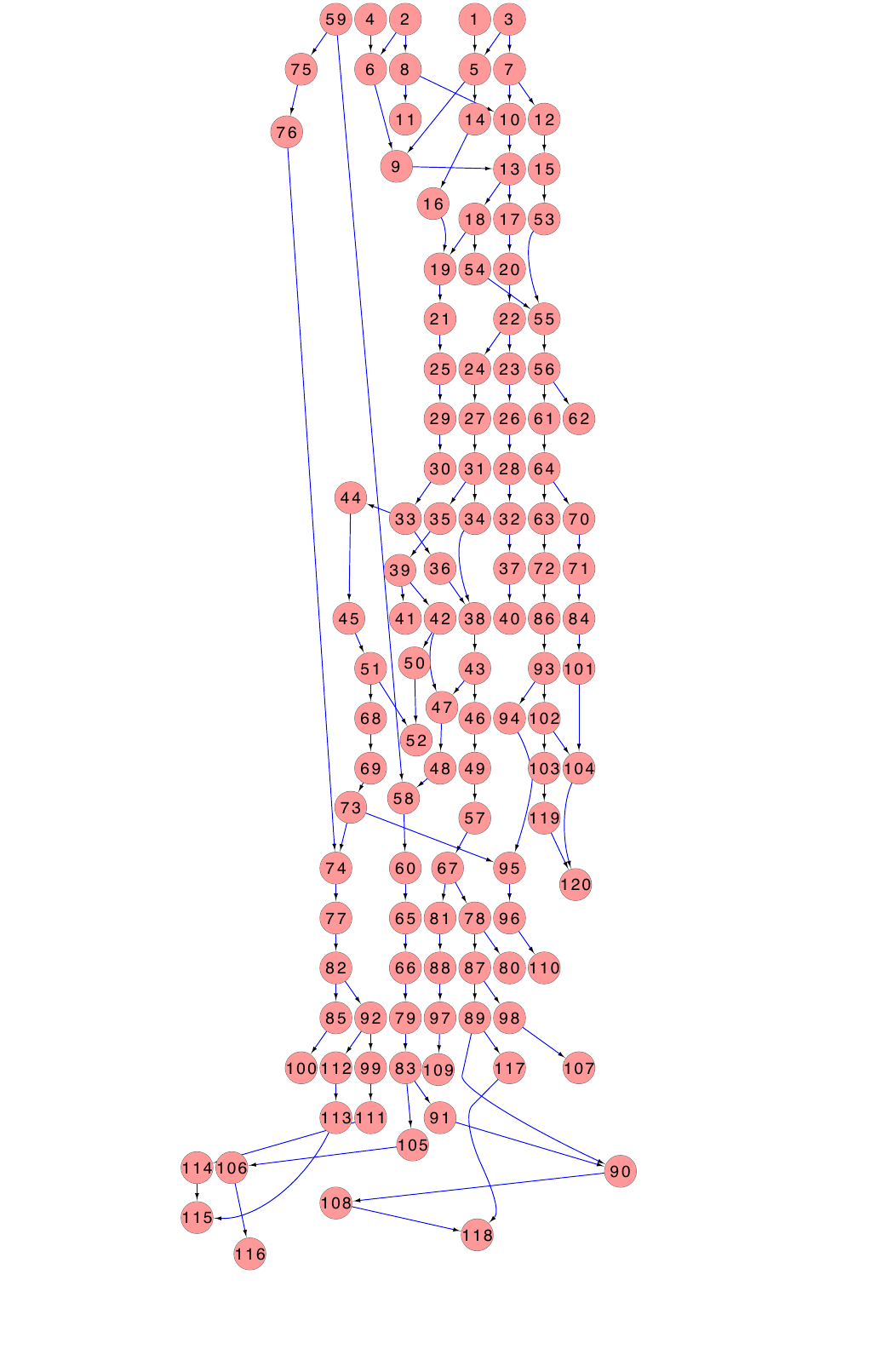}
\caption{Hierarchical representation of the 120 nodes network used to generate our simulated data. All the other smaller networks were subsets from this network}
\label{fi1}
\end{figure}

To start with we considered a real life biological network specifically the EGFR system. For this actual network we simulated a discrete dataset. This real life network consisted of only 14 nodes. In order to check efficiency of \textit{Graph\_sampler} for larger networks, we simulated networks of 5 to 120 nodes, with 100 data points for each node. Figure \ref{fi1} represents the network with 120 nodes. It is clearly a descending tree network. We used three different seeds to run three chains for each software program. We saved the three chains separately and calculated Gelman's $\hat{R}$ convergence diagnostic at each iteration. The first iteration for which $\hat{R}$ attained at most 1.05 for all edges of the graph was considered as the minimum number of iterations required for convergence. \textit{Graph\_sampler} was compiled with gcc version 4.2.1 (Apple Inc. build 5666) while \textit{structmcmc} was run with R 3.0.2 \citep{R3}.
We performed a time and convergence comparison between both the software. In order to check the performance and efficiency of both the software, we use R language script to plot the heat map and the accuracy curve. The heat map is used to summarize for all edges the edge posterior distribution through its means. We used the R language package \textit{lattice} to plot the heat maps. The accuracy curve is given by $accuracy$ $=$ $(true$ $positive$ $edges$ $+$ $true$ $negative$ $edges)/$ $total$ $number$ $of$ $edges$ and is a function of the probability threshold above which an edge is declared present. We used \textit{SDMTools}, to plot the accuracy curves.

\section{Results}
\subsection{\textbf{Dealing with Discrete data}}

\subsubsection{An actual biological network}
 
\begin{figure}[H]
  \centering
\includegraphics[width=0.3\textwidth]{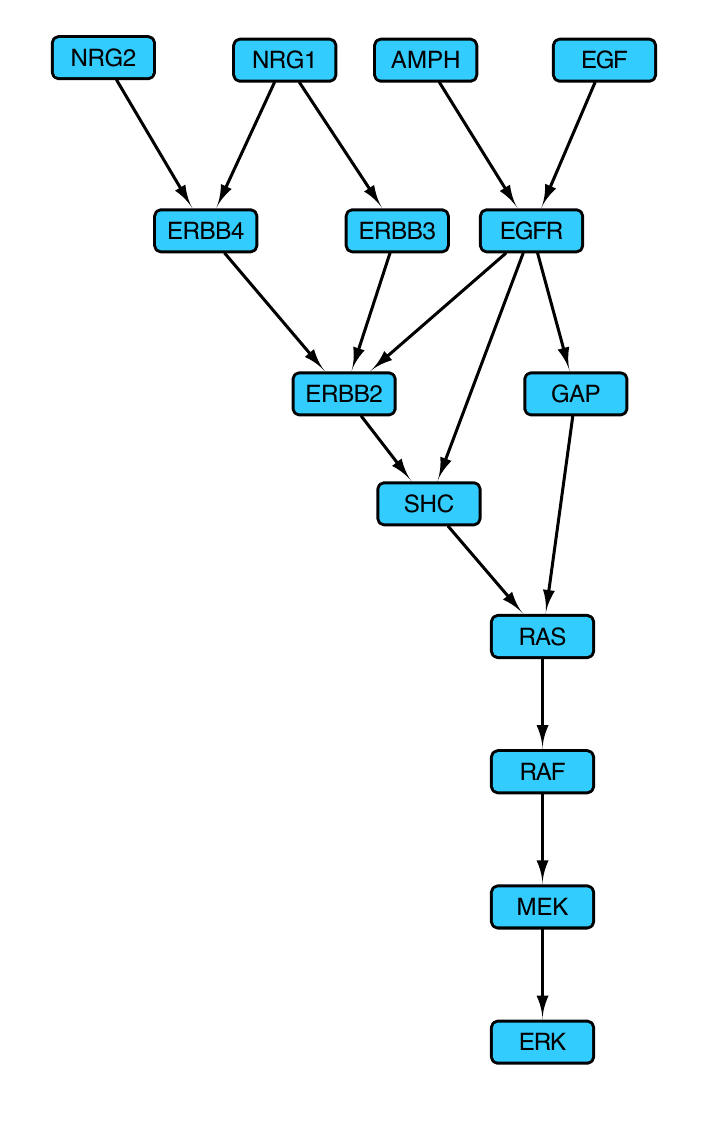}
\caption{Graph of the EGFR system}
\label{figcom6}
\end{figure}

To study the practical efficiency of \textit{Graph\_sampler} for a true biological network, we considered the same network as studied by S. Mukherjee \citep{Muk08}. In their work they considered a biological network known as the epidermal growth factor receptor (EGFR) system. Figure \ref{figcom6} gives a pictorial representation of the EGFR system. This biological network involves 14 proteins each of which is a ligand, a receptor or a cytosolic protein. Data for this study was synthesized based on the network and following the model as used by S. Mukherjee \citep{Muk08}. Each of the 14 proteins are considered to be a random variable with binary values $\left\{0, 1\right\}$. Depending on the parents, the conditional distributions are defined as Bernoulli with success parameter \textit{p}. The global parents are sampled with \textit{p}~=~0.5. For the child nodes, we take \textit{p} = 0.8 if one of its parents take the value 1 and \textit{p} = 0.2 otherwise. For each node we simulate 200 data points. 

To make our experimental runs coherent with \textit{structmcmc}, we defined the prior matrix on the graph structure based on the concordance prior proposed in \citep{Muk08}. As the model proposed was Multinomial model, we defined a Dirichlet prior for the parameters involved. In case of \textit{Graph\_sampler} we strengthen the prior on the structure of the network by considering a degree prior with $\gamma = 3$.

\begin{figure}[H]
  \centering
\includegraphics[width=1.0\textwidth]{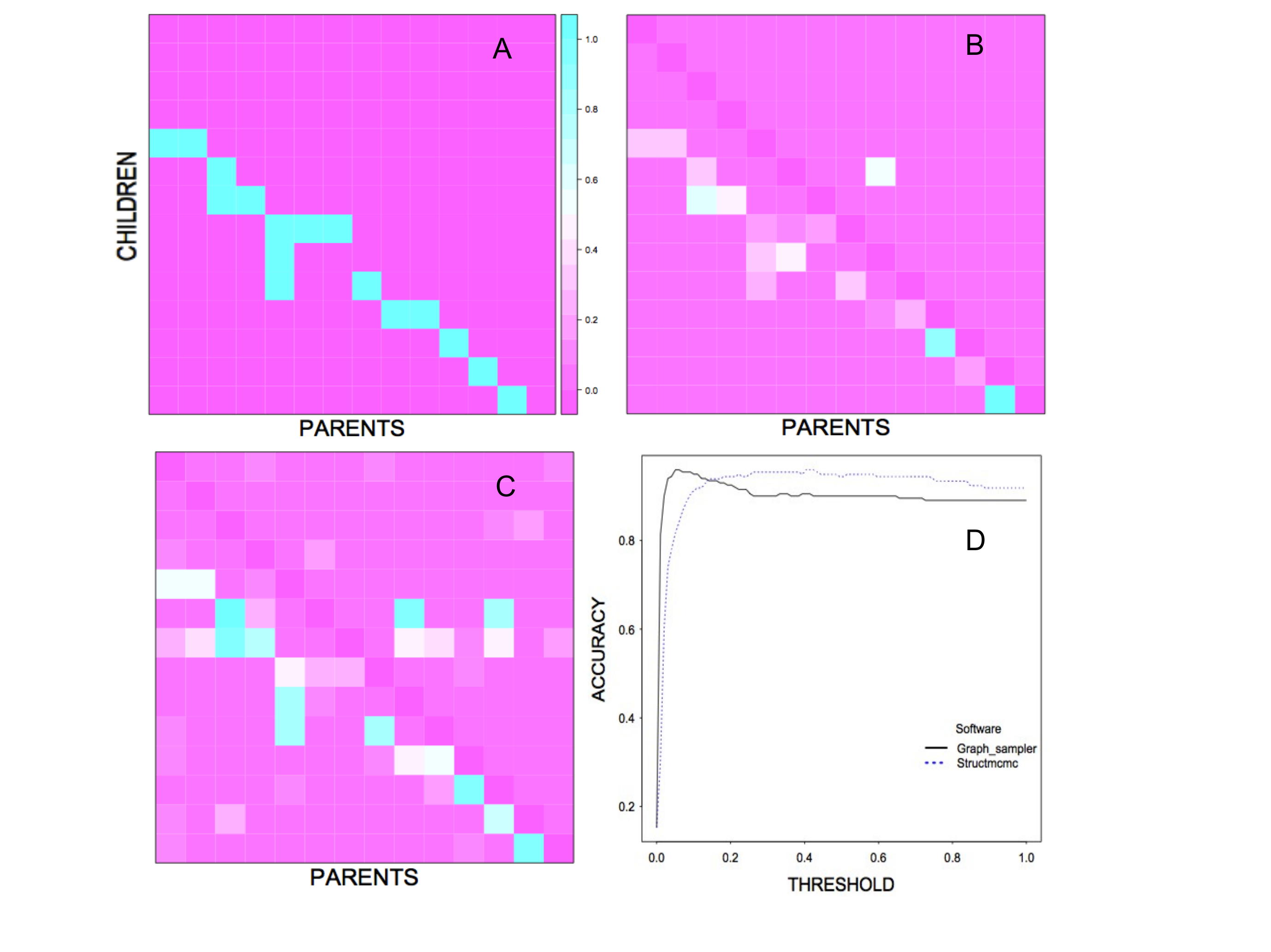}
\caption{Heat map of the real network with 14 nodes (A), edge posterior heat maps of $Graph\_sampler$ (B) $structmcmc$ (C) with discrete data. The x-axis represents the parent nodes and the y-axis the corresponding children. Accuracy curve of the two software (D).}
\label{figcom7}
\end{figure}

Figure \ref{figcom7} represent our comparative study in the form of heat maps and accuracy curve. We observe that for a small network like the EGFR system there is no significant difference in accuracy between \textit{Graph\_sampler} and \textit{structmcmc} for low threshold values but for higher threshold values, \textit{structmcmc} has an advantage. Observing the heat maps we find that even though \textit{Graph\_sampler} produces lower edge probability matrix but does not give rise to false negative edges. On the other hand \textit{structmcmc} generates high edge probability matrix but leads to a higher number of false negative edges that were not present in the prior network. Thus we can take this as a trade off where we have to be careful while working with both of these software and decide accordingly. If we focus on time efficiency of the two software, we observe that \textit{Graph\_sampler} is almost 200 times faster than \textit{structmcmc}. Thus taking both the criteria together, we observe that \textit{Graph\_sampler} has its own advantages and limitations like \textit{structmcmc}.
 
\begin{figure}[H]
  \centering
\includegraphics[width=1.0\textwidth]{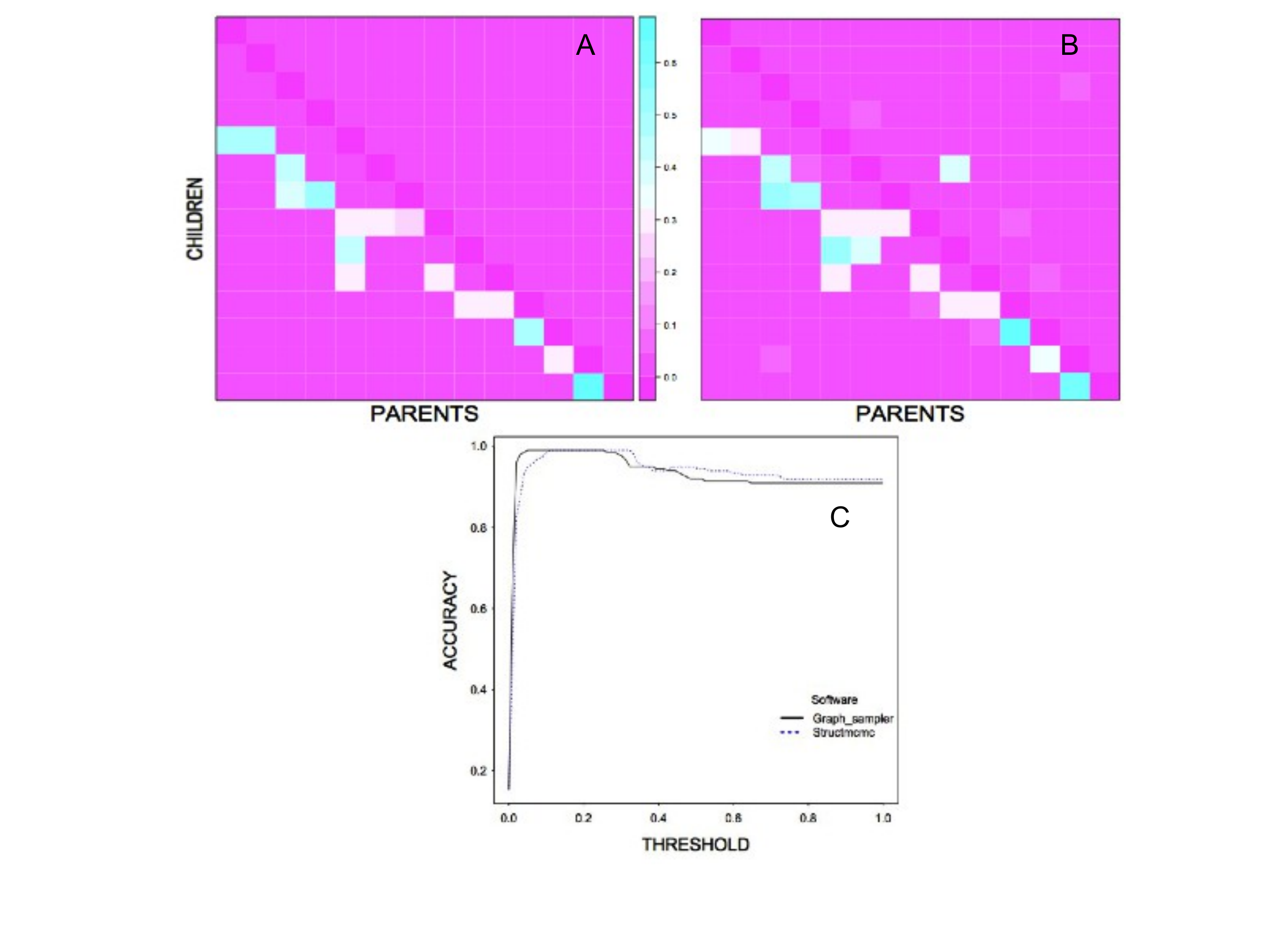}
\caption{edge posterior heat maps of $Graph\_sampler$ (A), \textit{structmcmc} (B) with Zellner g-prior, here the x-axis represents the parents and y-axis the corresponding children. Accuracy curve for the two software (C)}
\label{figcom8}
\end{figure}

An alternative other than the Dirichlet prior on the parameter could be the Zellner g-prior with g-prior equals 1. Even though the Zellner g-prior should be used for continuous data, we did use it for the discrete data to check its efficiency. For a binary data set as described above, we can easily fit a linear regression model and thus fall back to the continuous scenario making the use of Zellner prior valid. We still used the priors $P_C$ and $P_D$. Figure \ref{figcom8} represents the heat map and the accuracy curve for the two software. Comparing the heat map of figure \ref{figcom7}(B) and figure \ref{figcom8}(A), we observe that there is an improvement in the posterior edge probability matrix obtained from \textit{Graph\_sampler}. There is also a reduction in the number of false negative edges and this is true even for \textit{structmcmc}. As far as accuracy is concerned, both the software are quite efficient and there was no significant difference in accuracy between the two. However considering the time scale, \textit{Graph\_sampler} is again 100 times after than \textit{structmcmc} and thus have a slight advantage.

\subsubsection{Simulated networks}
\begin{figure}[!h]
\centering
\includegraphics[width=0.7\textwidth]{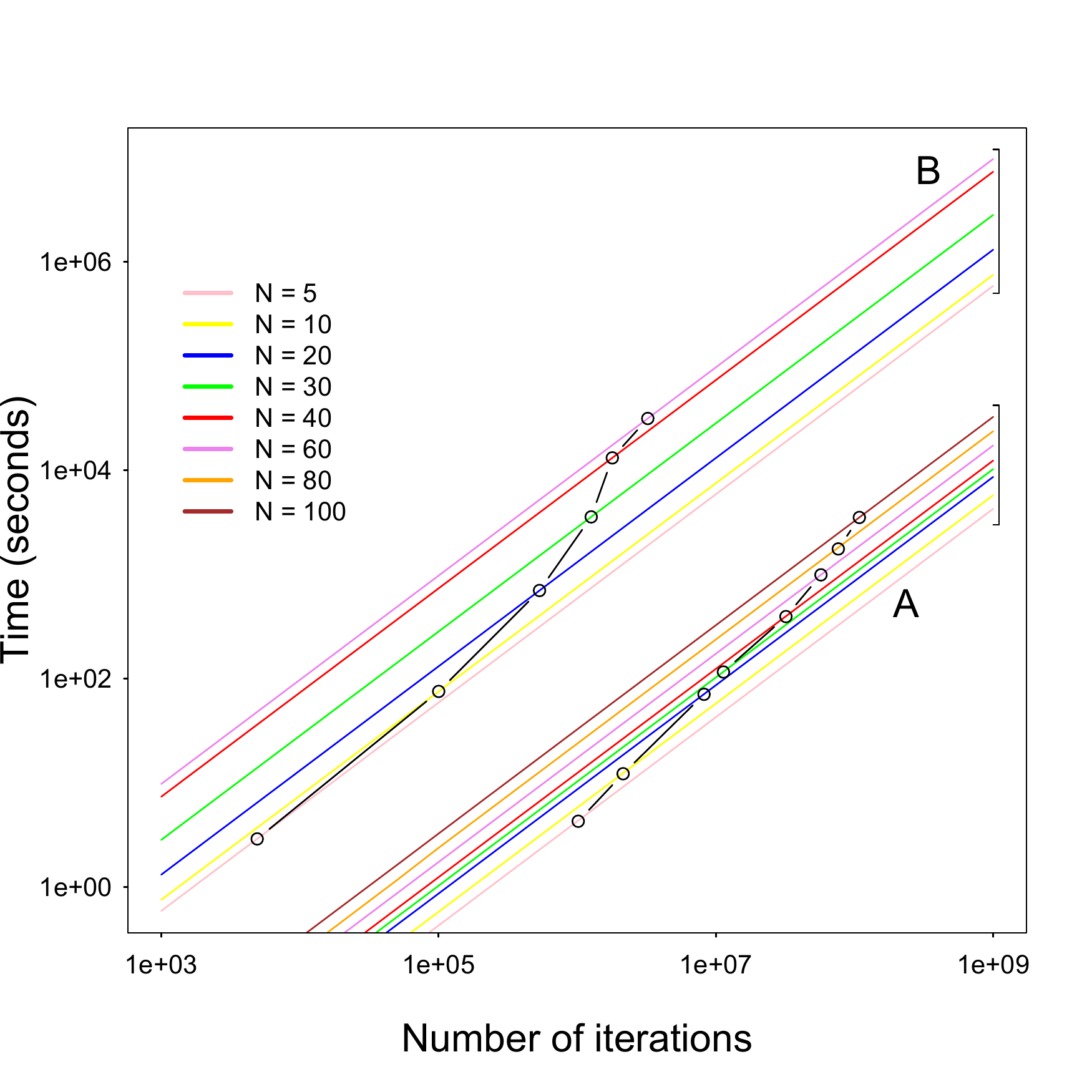}
\caption{Time and convergence comparison of \textit{Graph\_sampler} (A) and \textit{structmcmc} (B) performance for various network size (N) with discrete data. The x-axis represents the number of iterations performed and the y-axis the time taken on an Apple mac book 2.53 GHz Intel Core 2 Duo processor. The black lines with circles give the minimum number of iteration required to reach convergence.}
\label{figcom1}
\end{figure}

We studied the performance of \textit{Graph\_sampler} for discrete datasets, using the Multinomial model. Figure \ref{figcom1} gives a graphical summary of our timing results. Because of memory problems we could not achieve convergence with \textit{structmcmc} for networks of more than 60 nodes. Similarly for 120 nodes, \textit{Graph\_sampler} did not converge with a billion iterations. One of the reason behind this could be that with larger network size we have to increase the dataset. \textit{Graph\_sampler} was about 100 times faster than the \textit{structmcmc} for the same number of iterations. \textit{Graph\_sampler} running time was also less influenced by the network size (i.e. it increased by a factor 4.06 when going from 5 to 60 nodes with \textit{Graph\_sampler} and that factor being 16.5 for \textit{structmcmc}). With 30 nodes, \textit{Graph\_sampler} took about $2\times10^7$ iterations (275 seconds) to converge while for \textit{structmcmc} it was $10^6$ iterations (3848 seconds). Thus \textit{structmcmc} required 10 to 100 times less iterations to reach convergence but was about 14 times slower than \textit{Graph\_sampler} . We compared the edge probability matrices from both the software by the accuracy curve to check whether they converge to the true graph. 

\begin{figure}[!h]
 \centering
\includegraphics[width=0.9\textwidth]{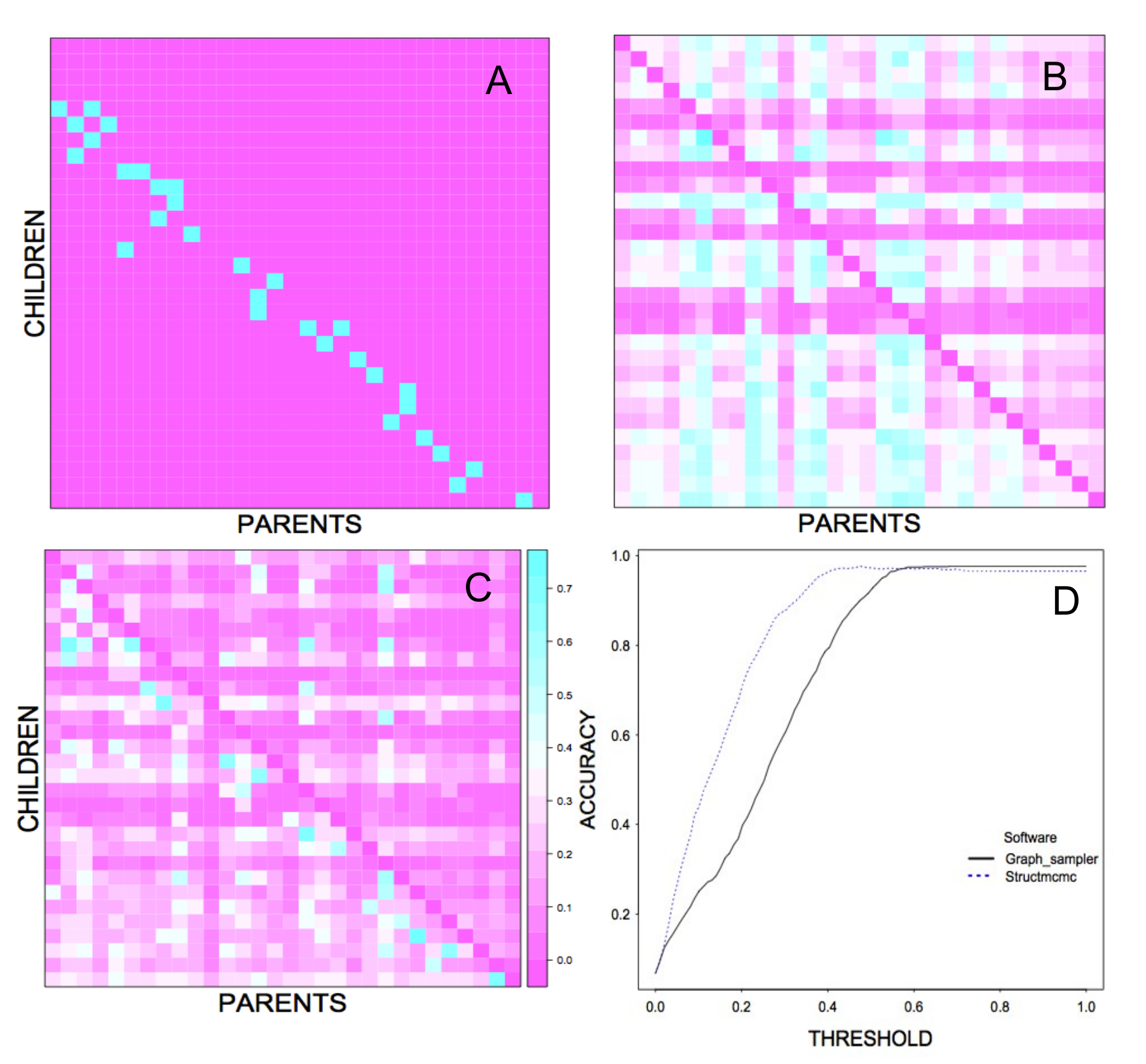}
\caption{Heat map of the true network with 30 nodes (A), edge posterior heat maps of $Graph\_sampler$ (B) $structmcmc$ (C) with discrete data. The x-axis represents the parent nodes and the y-axis the corresponding children. Accuracy curve of the two software (D).}
\label{figcom2}
\end{figure}

Figure \ref{figcom2} panels (A - C) represent the posterior edge probabilities in the form of heat maps. Figure \ref{figcom2} panel (D) shows the accuracy of the software in retrieving actual edges as a function of the probability threshold above which an edge is declared present. 
It was observed that with small threshold values \textit{structmcmc} had higher accuracy than \textit{Graph\_sampler}. We observe that with the single use of concordance prior the accuracy of \textit{Graph\_sampler} in retrieving edges correctly is low for small threshold values. However with a threshold value above 0.5, both the software had almost the same accuracy. Altogether, \textit{Graph\_sampler} reaches convergence faster in time and has equal accuracy as \textit{structmcmc} for threshold values above 0.5.

\begin{figure}[!h]
  \centering
\includegraphics[width=0.9\textwidth]{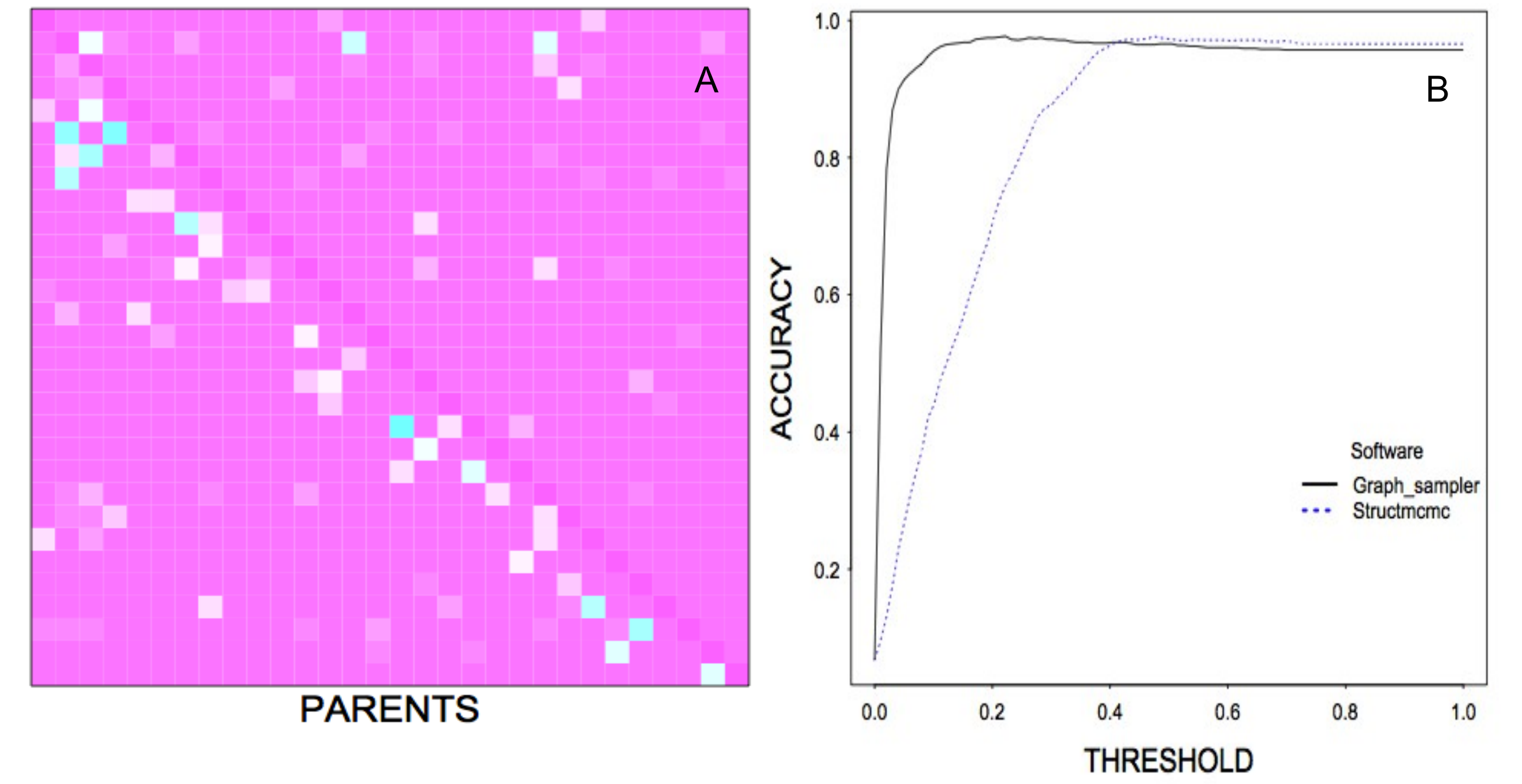}
\caption{Heat map of the posterior edge probabilities obtained from $Graph\_sampler$ with discrete data with concordance and degree priors (A) and the improvement in the accuracy curve due to the use of degree prior in \textit{Graph\_sampler} (B).}
\label{figcom21}
\end{figure}

The accuracy of \textit{Graph\_sampler} can be improved with the introduction of the degree prior $(P_D)$ with $\gamma = 2$. As our network is a descending tree network, the inclusion of the degree prior is very beneficial. Figure \ref{figcom21}(A) represents the posterior probabilities of the edges and resembles more like the true graph. It was observed that with the prior $P_D$, the accuracy of \textit{Graph\_sampler} improved significantly (Figure \ref{figcom21}(B)) and even with very low threshold values, the accuracy was almost equal to 0.9.

\subsection{\textbf{Dealing with Continuous data}}
For continuous data, \textit{structmcmc} uses a Zellner g-prior on regression parameters. With \textit{Graph\_sampler} either a normal-gamma prior or a Zellner g-prior can be used. The two models are compared here.
\subsubsection{Normal-gamma model}
\begin{figure}[!h]
\centering
\includegraphics[width=0.8\linewidth]{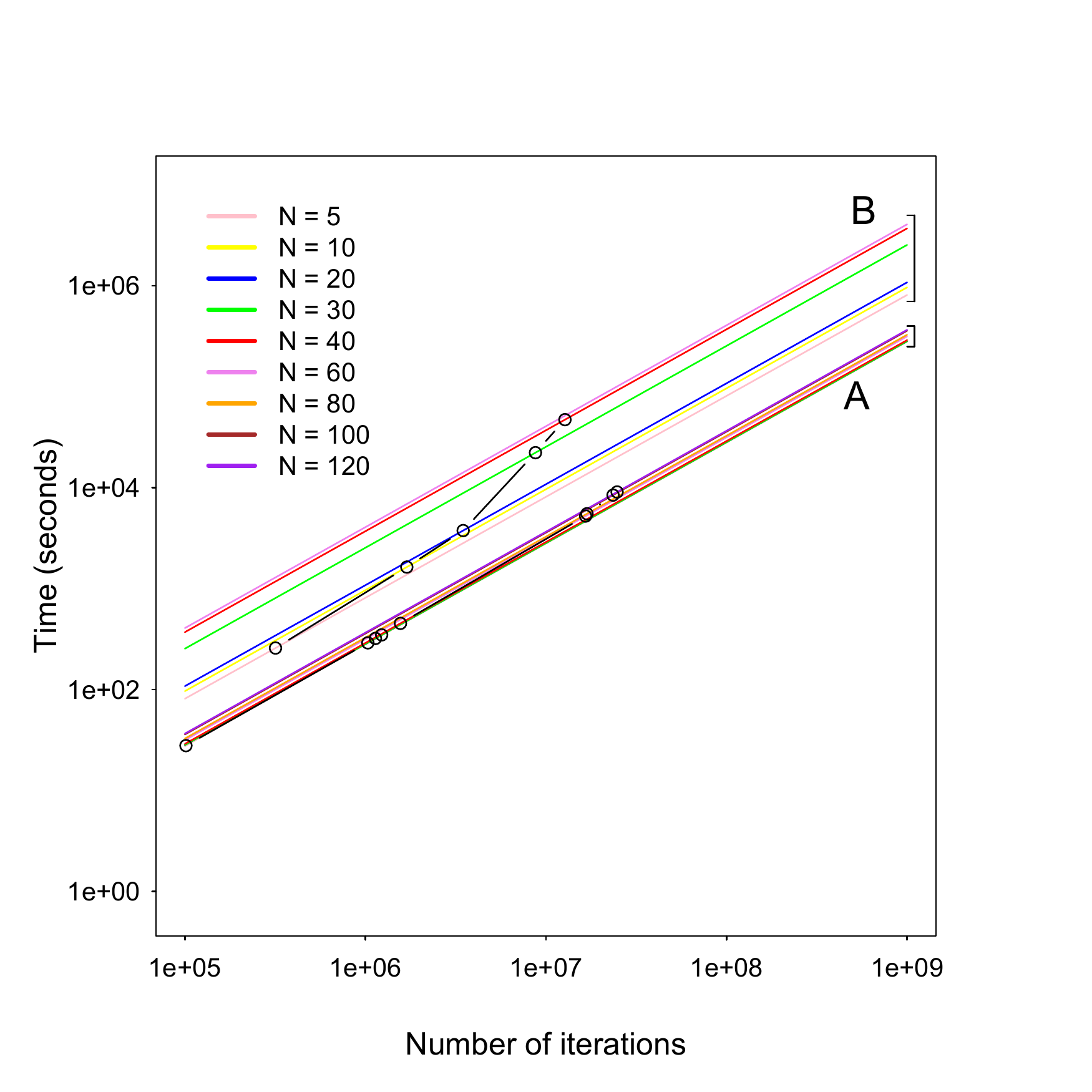}
\caption{Time and convergence comparison of \textit{Graph\_sampler}(A) with the \textit{structmcmc}(B) for varying network size with continuous data set. The \textit{Graph\_sampler} uses a normal gamma prior while \textit{structmcmc} uses a Zellner g-prior}
\label{figcom3}
\end{figure}

For the continuous data set, convergence with \textit{Graph\_sampler} was achieved with almost 10 times less iteration number compared to \textit{structmcmc}. It was observed that with networks having 60 nodes, \textit{structmcmc} faced problems in convergence. Figure \ref{figcom3} represents our study in a graphical way. Regarding the time taken for iterations, \textit{Graph\_sampler} was almost 10 times faster than \textit{structmcmc}. For networks with more than 20 nodes, time efficiency of \textit{structmcmc} decreases sharply. The very narrow band width (A) reveals that for \textit{Graph\_sampler} the increase in network size does not have much influence on time. Figure \ref{figcom3} represents our study in a graphical way.     

\begin{figure}[!h]
  \centering
\includegraphics[width=0.9\textwidth]{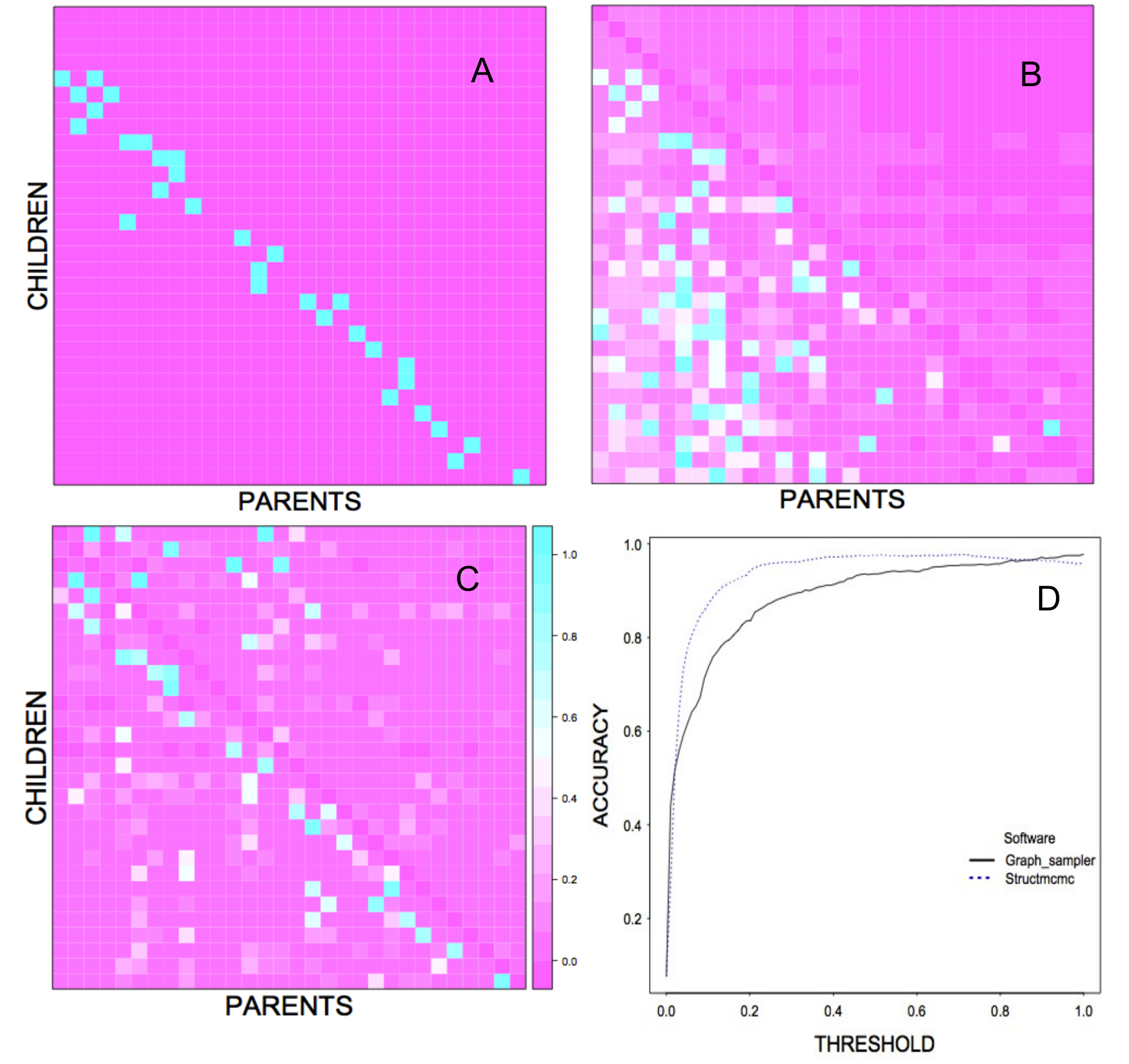}
\caption{Heat map of the true network with 30 nodes (A), edge posterior heat maps of $Graph\_sampler$ (B) $structmcmc$ (C) with continuous data, the x-axis represents the parents and y-axis the corresponding children. Accuracy curve for the two software (D).}
\label{figcom4}
\end{figure}

We also studied the posterior edge probabilities obtained from each of the software to draw inference on their efficiency to retrieve the true graph structure. We plotted the posterior edge probabilities in the form of heat maps to understand sampling scheme prescribed in \textit{Graph\_sampler}. Figure \ref{figcom4} represents the three heat maps for a network with 30 nodes. We observe that \textit{Graph\_sampler} perform well in retrieving edges present higher up in a tree network. However as we go down the tree structure, the efficiency of \textit{Graph\_sampler} decreases. As our model used simulated data, so there is an underlying correlation in the dataset. For this reason as we go down the network structure \textit{Graph\_sampler} samples many new edges. Interestingly the directionality of the network structure is maintained. On the other hand \textit{structmcmc} is efficient as we move down the network structure. However \textit{structmcmc} fumble with the directionality of the edges. We plotted the accuracy curve for the two software. Figure \ref{figcom4}(D) represents that the accuracy of \textit{structmcmc} is slightly higher than that of \textit{Graph\_sampler} for smaller threshold values. However this difference is not much and for higher threshold values, both the software have almost equal accuracy.

\subsubsection{Strengthening the prior model by considering the hierarchy of the structure and defining a degree prior }
\begin{figure}[!h]
  \centering
\includegraphics[width=0.85\textwidth]{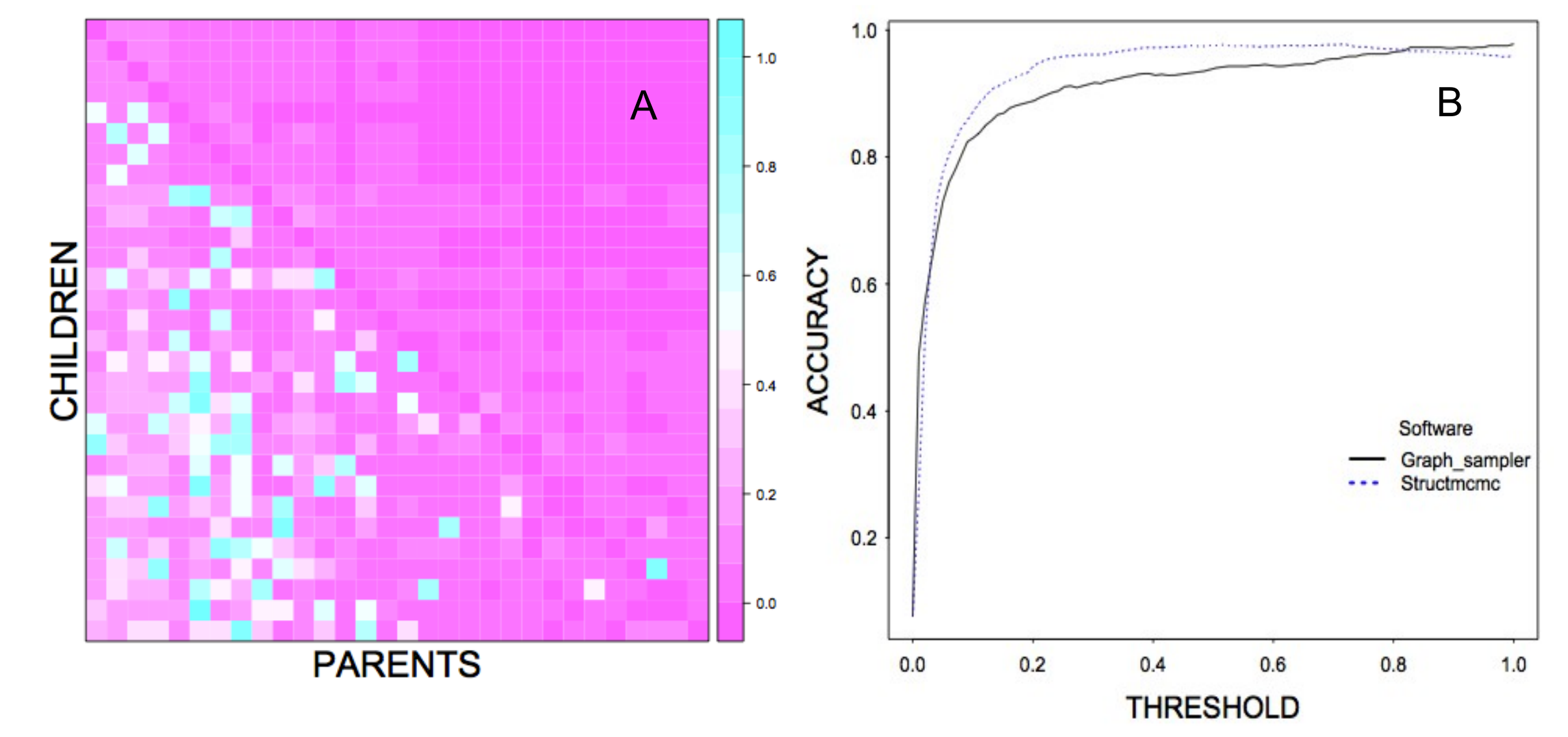}
\caption{Heat map of the posterior edge probabilities obtained from $Graph\_sampler$ with Normal-gamma model with concordance and degree priors (A). Accuracy curve with the use of degree prior in \textit{Graph\_sampler} (B).}
\label{figcom41}
\end{figure}

The degree prior $(P_D)$ with $\gamma = 2$ was introduced to check the improvement in the accuracy of \textit{Graph\_sampler}.  Figure \ref{figcom41}(A) represents the posterior probabilities of the edges and resembles more like the graph in Figure \ref{figcom4}(B). It was observed that with the prior $P_D$, there was no significant improvement in the accuracy of \textit{Graph\_sampler} (Figure \ref{figcom41}(B)).

\subsubsection{An alternative: Zellner g-prior model}

For continuous datasets, we can also use the Zellner g-prior as used in \textit{structmcmc}. Unlike \textit{structmcmc}, we run \textit{Graph\_sampler} for various g-prior values. We started with a g-prior of 1 for all the networks considered in our study. We checked the time required and the convergence point. Later we performed similar runs for different g-prior with values 5, 10, 50 and 100. In each case we observed the convergence rate and the posterior edge probabilities. 

With a g-prior value of 1 or 5, convergence was achieved for all the networks. As we increased the value of g-prior to 10, runs with networks having 5 and 10 nodes converged with the maximum value of $\hat{R}$ being 1.12. This was not the case for networks with 20, 30, 40 and 60 nodes as they converged ($\hat{R}$ = 1.05 approx). However increasing the g-prior value to 50 and 100, convergence was not achieved for any of the networks. Thus in such a situation, \textit{Graph\_sampler} and \textit{structmcmc} differs. For smaller network sizes, \textit{Graph\_sampler} performed well with g-prior value equal to 1 or 5. With bigger networks sizes having 100 data points for each nodes, the g-prior value can range from 1 to 40. \textit{Graph\_sampler} failed to converge when the g-prior value was equal to the number of data points. On the other hand, \textit{structmcmc} performed well with higher g-prior values and was most efficient when the g-prior value was equal to the number of data points.

In order to understand the reason behind such a difference in convergence rate between the two software, we discuss the flipping technique used in \textit{Graph\_sampler} along with its advantages and disadvantages.

\subsubsection{Problem with flips}

The primary advantage of the reduced jump kernel (only adding or deleting one edge at a time) used in \textit{Graph\_sapmpler} is that it is faster than the jump kernel allowing flips. Since the choice of pairs of nodes is systematic, there is no need to check the neighbourhood cardinality \citep{Hus03} as done in \textit{structmcmc}. 

This jump kernel has some drawbacks also. Consider a network with 5 nodes where node 4 is a parent of node 5. In the MCMC simulations, at a particular step, we propose to add an edge from node 5 to node 4. As the log posterior for the proposed network is quite high (if 4 conditions 5, the two are correlated and 5 conditioning 4 has high probability), we accept such a proposal. According to our flipping technique, in order to retrieve the true edge (from 4 to 5), we need to first delete the edge from 5 to 4, leaving them independent. However a network with 4 and 5 independent has low log posterior probability and we rarely accept such a move.

\begin{figure}[!h]
\centering
\includegraphics[width=0.7\linewidth]{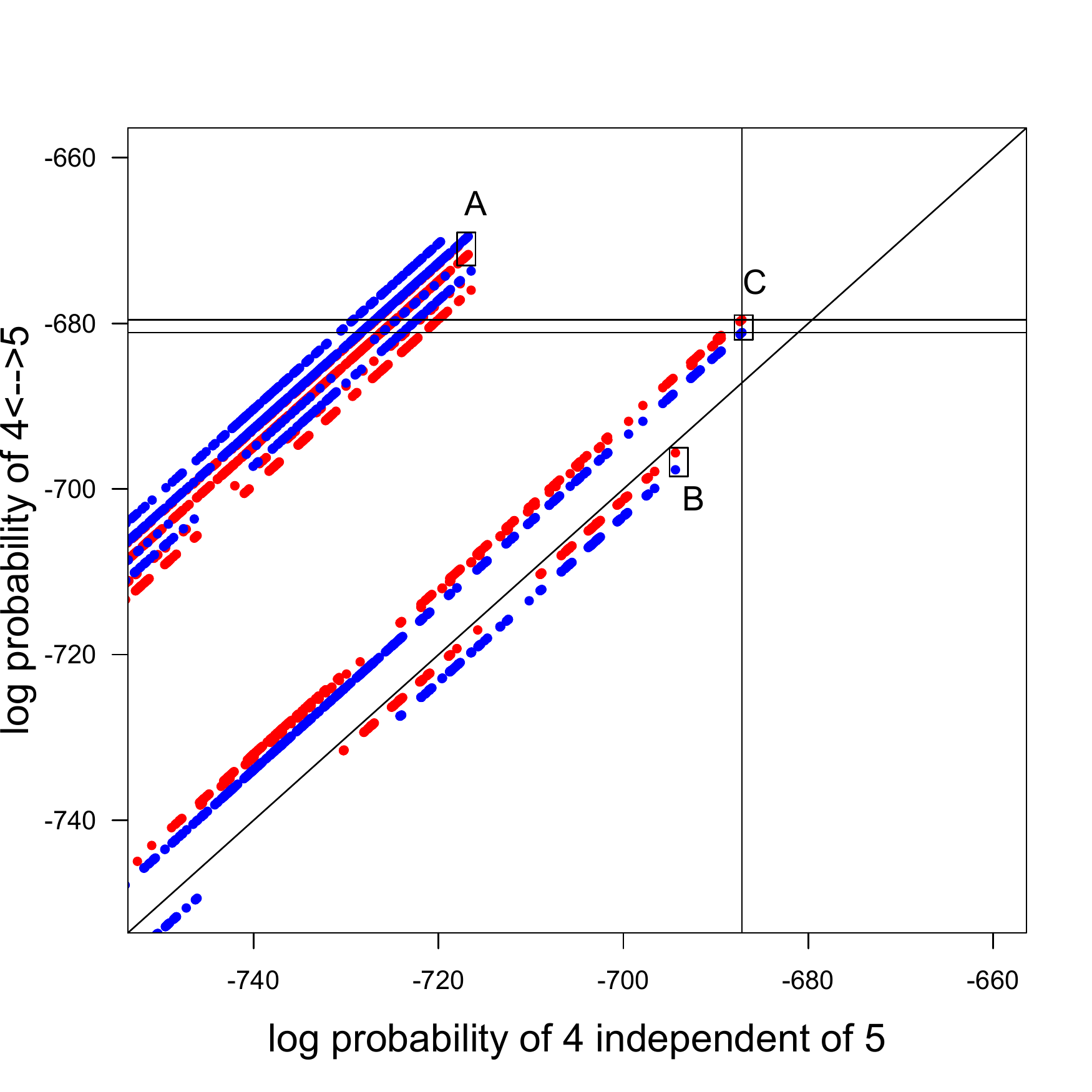}
\caption{The flips for a network with 5 nodes along with the log posterior probability after each flip. The blue dots represent the presence of the edge from 4 to 5 and the red dots for the edge from 5 to 4 in a 5 node network. The vertical and the horizontal lines in the graph are kept for easy reference of the marked pairs with both the axis.}
\label{fig7}
\end{figure}

Figure \ref{fig7} is a dot plot where the blue dots and the red circles appearing in pairs represents the difference in log probability for a network when passing from an edge (4 to 5) to an edge (5 to 4) respectively using the jump kernel specified in \textit{Graph\_sampler}. For a pair A, a move from the red dot (log probability -675) to a blue dot (log probability -676), the log probability has to pass through -720 (4 and 5 independent) making such a move impossible. So in the most likely regions of $G$ the flip is very unlikely. For the pair B, a move from the red dot (log probability of -698) to a state of independence (log probability of -685) is easy, but the next move to a blue dot (log probability of -700) is not easy. This is also an unlikely region where some flips are possible. The flip for the pair C is unlikely to occur as the log probability has to pass by -687 while going from -678 to -681. Flips would occur easily when they are close to the diagonal.

The difficulty with the jump kernel can be due to the large data set for which the posterior mass favours fewer graphs. Under such a situation, the standard MCMC scheme faces difficulty in moving between graphs, or finding the high-scoring graphs. In such a case parallel tempering is a proficient option to speed up the MCMC-based convergence of network inference. This tempering approach is generally referred to as Model Composition by Metropolis-Coupled Markov Chain Monte Carlo (MC\textsuperscript{4}) \citep{Sach10}. The parallel tempering involved in this MC\textsuperscript{4} (beyond the scope of this paper) approach allows proper mixing of the Markov chain and helps to escape the local maxima. 

\subsubsection{Sensitivity to the prior on graph structure}
To check the sensitivity of the normal-gamma model with respect to informative and non informative priors, we considered a network with 40 nodes having 100 data points for each node and defined only the $P_B$ priors on the edges. We first considered an informative prior on edges with each desired edge having a prior probability of 0.9 and 0.1 for others except for autoloops for which probability was 0. We define a less informative prior with 0.8 and 0.2 and carry out our experimental run.  We repeated the process and finally defined a flat non informative prior of 0.5 for all the edges. 

For each run with different prior probabilities, convergence was obtained in \textit{Graph\_sampler} and then retrieved the posterior edge probabilities as it conveyed the information regarding the sensitivity of \textit{Graph\_sampler} in selecting the desired graph out of all the equivalent graphs.

\begin{figure}[!h]
  \centering
\includegraphics[width=0.65\textwidth]{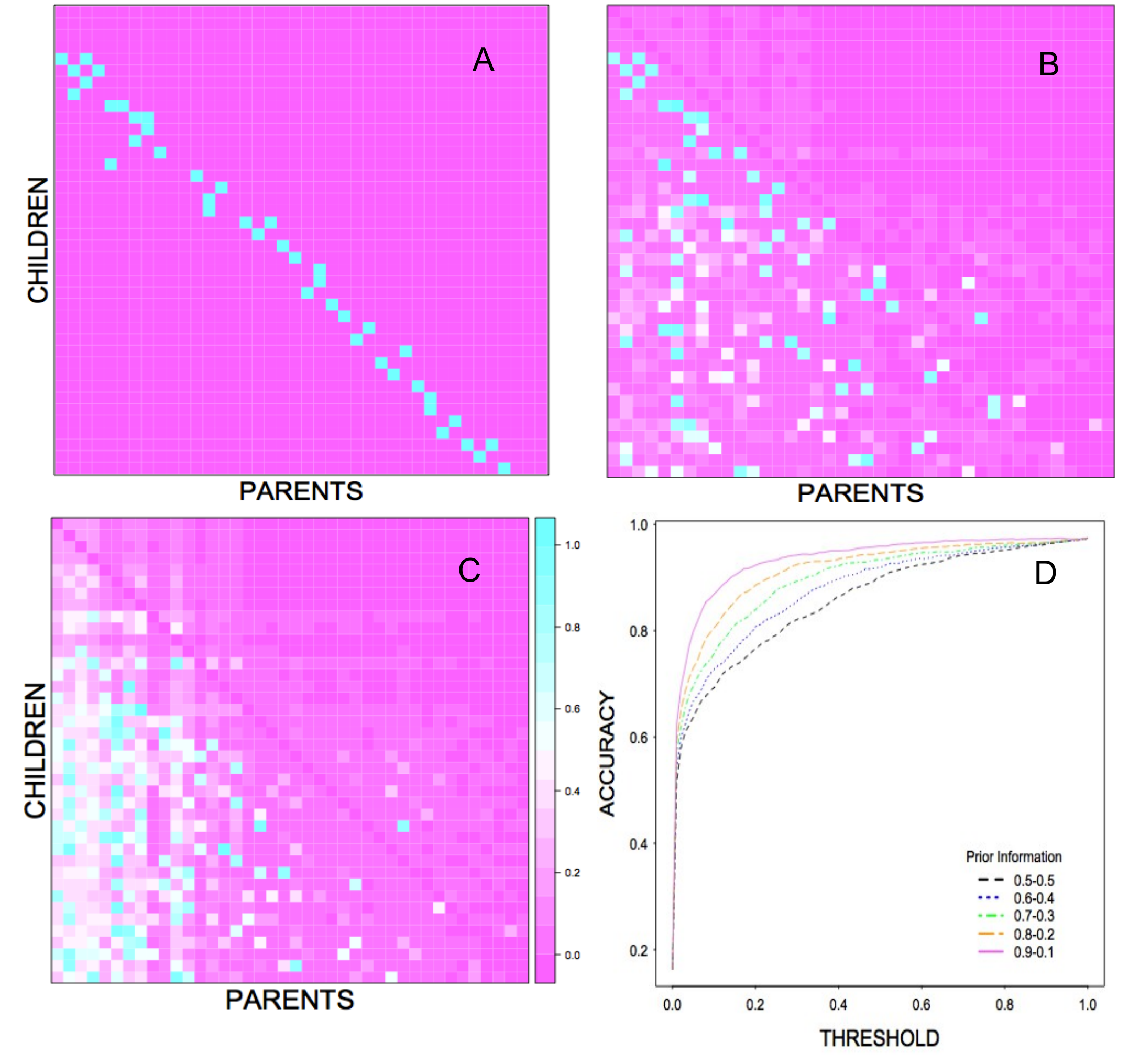}
\caption{Heat map of the true (desired) network with 40 nodes (A), heat map of posterior edge probabilities obtained from $Graph\_sampler$ with an informative prior of 0.9 for desired edges (B) and with a noninformative prior of 0.5 (C) with a Normal gamma likelihood; the x-axis represents the parents and y-axis the corresponding children. Accuracy curve for the various informative priors (D)}
\label{fig5}
\end{figure}

With a strong informative prior of 0.9 and 0.1, \textit{Graph\_sampler} converged and was able to retrieve all the desired edges with a high probability. The posterior probabilities of some undesired edges were also high. This is mainly observed as we move down the network due to the presence of partial correlation between the nodes higher up in the network and those at the bottom. As we use less informative priors, this behaviour becomes more prominent and the efficiency of \textit{Graph\_sampler} decreases (Figure \ref{fig5}). With a flat prior of 0.5 for all the edges, the efficiency of \textit{Graph\_sampler} is the least. Figure \ref{fi1} and the heat map of Figure \ref{fig5}(A) show that the true network is a descending tree (the upper triangular matrix of the heat map has zero edge probability). For the informative prior, we observe that the normal-gamma model works well for the upper part of the tree network. However as we descend down the tree, the sensitivity of the model decreases as more children are involved. One way to increase the performance of the model can be by increasing the number of data points for each node present in the lower part of the network. We plotted the accuracy curve to depict the efficiency of \textit{Graph\_sampler} for the various informative priors.

Figure \ref{fig5}(D) represents the accuracy curve of \textit{Graph\_sampler}. We observed that \textit{Graph\_sampler} was very versatile with the type of prior information provided. With very strong prior the accuracy was almost equal to 0.9 for threshold values above 0.2. This stated that \textit{Graph\_sampler} was efficient enough to retrieve the true edges with higher posterior probabilies and allot low probability (less than the threshold) to false edges. For noninformative priors \textit{Graph\_sampler} had an accuracy of 0.65 for threshold below 0.3. The accuracy increased to 0.8 and above with the increase in threshold from 0.5 to 1.0.

\begin{figure}[H]
  \centering
\includegraphics[width=0.7\textwidth]{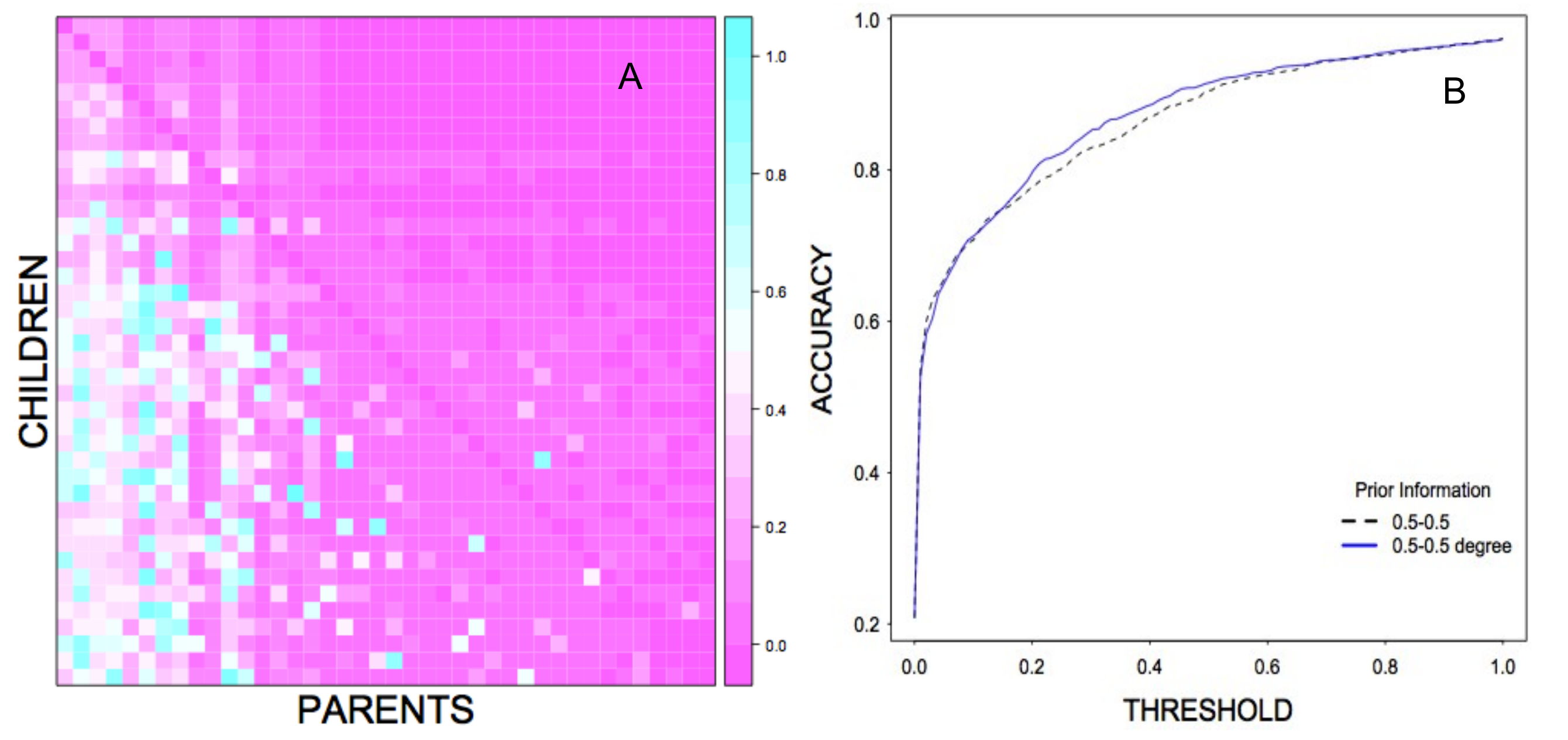}
\caption{Heat map of the posterior edge probabilities obtained from $Graph\_sampler$ with Normal-gamma model with a noninformative Bernoulli prior and degree priors (A). Accuracy curve due to the use of degree prior in \textit{Graph\_sampler} (B).}
\label{figcom51}
\end{figure}

We checked the sensitivity of \textit{Graph\_sampler} for degree prior $(P_D)$ with a noninformative Bernoulli prior on edges. We observed that there was not much improvement in the~accuracy~of~\textit{Graph\_sampler}~in~this~case~(Figure~\ref{figcom51}). 

\section{Discussion}

We observed that like other widely used software (\textit{bnlearn, deal}), \textit{Graph\_sampler} works well with both discrete as well as continuous data. For continuous data, we considered a regression model and we fitted a multinomial model for the discrete data. These two model representations are commonly used when dealing with BNs. For continuous data, \textit{Graph\_sampler} allow the user to use either the Normal-gamma prior or the Zellner g-prior for the parameters involved in the regression model. Our results showed that \textit{Graph\_sampler} performs better with the Normal-gamma prior than the Zellner g-prior. Another drawback of using a Zellner g-prior is that while using this prior the number of parents should be less than the number of data points per node which can be unrealistic while dealing with real observations. For the multinomial model we defined a Dirichlet prior for the parameters. This software provides a number of different structure priors which are mutually independent and can be used in varied combination to improve the results. In particular we considered the concordance prior ($P_C$), the Bernoulli prior ($P_B$), the degree prior ($P_D$) and the motif prior ($P_M$). Even though we did not utilize the motif prior in this work, but one can use this prior according to the needs. Taking into consideration the data type and the structure prior we are able to perform full Bayesian analyses to sample graphs from the specified posterior distribution. We observed that while dealing with large networks the choice of priors and their combination is quite essential for retrieving the true edges because of the huge number of possible DAGs. It means that larger networks generally require strong informative priors on the structure of the network. Hence for specific network structures (like, tree structure), we have to think of incorporating a more appropriate structure prior. For a large tree structure we observed that the efficiency of \textit{Graph\_sampler} in retrieving the true edges decreases as we move down the tree. An alternative way to solve this problem could be to use larger data sets for nodes present way down in the tree structure. 

We find that \textit{Graph\_sampler} is very efficient with respect to time. Both for continuous as well as discrete data, we observed that \textit{Graph\_sampler} was atleast 10 times faster than \textit{structmcmc}. One of the primary reason behind \textit{Graph\_sampler} being time efficient is due to its fast jumping kernel. The systematic scanning of the edges and proposing to add or delete an edge at each iteration makes this kernel very efficient with respect to time. Moreover this jumping kernel requires less memory for storage. However as explained in Section 5, this jumping kernel has some limitations. The kernel faces difficulty with large data sets for which the posterior mass favours fewer graphs. The standard MCMC scheme faces problem with the local maxima and thus fails to search for the high-scoring graphs. A proficient option would be to use parallel tempering (beyond the scope of this paper) to speed up the MCMC based convergence. Secondly, \textit{Graph\_sampler} is scripted in C language. C language being a compiled language is much more faster than R which is an interpreted language. 
Our results showed that for large networks i.e. networks with more than 100 nodes, \textit{Graph\_sampler} was efficient in convergence. To resume, we observed that \textit{Graph\_sampler} is a flexible software which is efficient with respect to time and convergence even for large networks.

We observed that \textit{Graph\_sampler} could be a good software apart from the widely used R packages to perform Bayesian analyses of DAGs models. Besides \textit{structmcmc}, \textit{bnlearn} and \textit{deal} which are scripted in R, there are several other very efficient BN software that are capable of performing Bayesian inference on parameter estimation and/or on network structure. Some of these software are free while others run on commercial platforms. Each software run on different platforms and follow specific sampling scheme. For interested readers, \cite{Kor10} reviewed some of the software available at the time that do inference on network structure.

\bibliographystyle{spbasic}
\bibliography{Paper}


%

\end{document}